\documentclass[notitlepage,twocolumn,showpacs,superscriptaddress,nofootinbib,preprintnumbers,aps,prl]{revtex4-1}  

\usepackage{graphicx}  
\usepackage{dcolumn}   
\usepackage{bm}        
\usepackage{amssymb}   
\usepackage{amsmath}
\usepackage{slashed}
\usepackage{mathtools}
\usepackage[unicode=true,
 bookmarks=true,bookmarksnumbered=false,bookmarksopen=false,
 breaklinks=false,pdfborder={0 0 1},backref=false,colorlinks=true]{hyperref}
\usepackage{cleveref}
\usepackage{xcolor}

\allowdisplaybreaks

\newcommand{\cA}{\mathcal{A}}

\newcommand{\nn}{\nonumber \\}

\def\f#1#2#3{f^{#1}_{\ #2#3}}

\begin{document}


\preprint{CALT-TH-2022-036}

\title{Geometry in Scattering Amplitudes}
\author{Andreas Helset}                             

\affiliation{Walter Burke Institute for Theoretical Physics, California Institute of Technology, Pasadena, CA 91125}

\author{Elizabeth E.~Jenkins}                             

\affiliation{Physics Department, University of California San Diego, 9500 Gilman Drive, La Jolla, CA 92093-0319}

\author{Aneesh V.~Manohar}                             

\affiliation{Physics Department, University of California San Diego, 9500 Gilman Drive, La Jolla, CA 92093-0319}

\date{\today}

\begin{abstract}
	We formulate the field-space geometry for an effective field theory of scalars and gauge bosons.
	Geometric invariants such as the field-space curvature enter in both scattering amplitudes and the renormalization group equations, with the scalar and gauge results unified in a single expression.
\end{abstract}

\maketitle


\section{\label{sec:intro}Introduction}

One central property of the $S$-matrix is its invariance under field redefinitions \cite{Chisholm:1961tha,Kamefuchi:1961sb,Arzt:1993gz}. The form of the Lagrangian may change under a field redefinition, but the scattering amplitudes---and physical observables---remain unchanged. For the case of non-derivative scalar field redefinitions, this invariance can be understood through the geometry of field space where the scalar fields are the coordinates. In this picture, a non-derivative scalar field redefinition is a coordinate change on the field-space manifold, which does not change the dynamics. The geometric approach was used to characterize deviations from Standard Model scattering amplitudes in terms of the curvature of the scalar manifold~\cite{Alonso:2015fsp,Alonso:2016oah,Cohen:2021ucp,Alonso:2021rac}.

Recently, the geometric underpinning of the invariance of scattering amplitudes under field redefinitions has been extended to include field redefinitions with derivatives and higher-spin fields through several approaches \cite{Cheung:2022vnd,Cohen:2022uuw}. However, these generalized geometric approaches have yet to explain how to make manifest the invariance of scattering amplitudes under field redefinitions including derivatives at loop level. 

The geometry of field space is also a practical tool. Many terms in a Feynman-diagram expansion are reorganized into curvature invariants, which allows for efficient calculations of scattering amplitudes.  The need for Feynman diagrams can be circumvented by using on-shell bootstrap methods. In this case the field-space geometry is useful, as it gives physical meaning to the unfixed coefficients in scattering amplitudes that are constructed using on-shell methods. The field-space geometry plays an important role in the soft limit for scalar effective field theories \cite{Cheung:2021yog}, in an effective action invariant under field redefinitions \cite{Vilkovisky:1984st}, and in the renormalization group equations for scalar effective field theories~\cite{Alonso:2015fsp,Alonso:2016oah}, including to higher loop order~\cite{Alonso:2022ffe,Jenkins:2023bls}.

In this paper, we summarize some pertinent properties of the geometry of scalar field space, and show how several of these properties extend to an effective theory of scalars and gauge bosons. Many concepts from differential geometry apply to theories with scalars and gauge bosons, e.g., field-space metric, covariant derivative, Riemann curvature, and Killing vectors.
We also find a geometric description of the scalars and gauge fields which extends to loop level; in particular, we derive renormalization group equations which only depend on geometric quantities.



\section{\label{sec:manifold}Field-Space Manifold}

We consider a theory of scalars and gauge bosons with interactions with at most two derivatives,\footnote{Higher-derivative interactions are linked to derivative field-redefinitions, which is outside the scope of this work. They have been considered in Refs.~\cite{Cheung:2022vnd,Cohen:2022uuw}} and  ignore CP-violating interactions for simplicity. We group the scalars and gauge bosons into real multiplets $\phi^{I}$ and $A^{B\mu_B}$, where $I,J,K,\dots$ are scalar indices and $(A\mu_{A}), (B\mu_{B}),\dots$ are gauge and Lorentz indices, treated as a combined index. The general gauge-invariant Lagrangian takes the form
\begin{align}
	\label{eq:Lagr}
	\mathcal{L} &= \frac{1}{2} h_{IJ}(\phi) (D_{\mu}\phi)^{I} (D^{\mu}\phi)^{J} - V(\phi)  \\ 
	    &- \frac{1}{4} g_{AB}(\phi) F^{A}_{\mu\nu} F^{B\mu\nu} , \nonumber 
\end{align}
where $h_{IJ}(\phi)$, $V(\phi)$, and $g_{AB}(\phi)$ depend on the scalar fields. The covariant derivative of the scalar field and the field strength are 
\begin{align}
	\label{eq:covD}
	(D_{\mu}\phi)^{I} &= \partial_{\mu} \phi^{I} + A^{B}_{\mu} t^{I}_{B}(\phi) , \\
	F^{B}_{\mu\nu} &= \partial_{\mu} A^{B}_{\nu} - \partial_{\nu} A^{B}_{\mu} - \f{B}{C}{D} A^{C}_{\mu} A^{D}_{\nu} ,
\end{align}
where $t^I_A(\phi)$ are Killing vectors of the scalar manifold. The Lie derivative of the scalar metric $h_{IJ}$ vanishes,
\begin{align}
	\label{eq:LieMetricH}
	t^{K}_{A} h_{IJ,K} + h_{KJ} t^{K}_{A,I} + h_{IK} t^{K}_{A,J} = 0 ,
\end{align}
where $h_{IJ,K} = \partial_{K} h_{IJ}$ and $t^{I}_{A,J} = \partial_{J} t^{I}_{A}$. 
The Killing vectors satisfy the Lie bracket relations
\begin{align}
[t_A,t_B]^I =& \f CAB t^I_C \,.
\end{align}
The gauge coupling constant is included in $t^I_A$, and hence also in the structure constants $\f CAB$.\footnote{In the Standard Model effective field theory (SMEFT), the vector $t^{I}_{A}$ is 
\begin{align}
	t^{I}_{A} = -\frac{1}{2} \tilde \gamma^{I}_{A,J} \phi^{J} ,
\end{align}
where the 4-by-4 matrices $\tilde\gamma^{I}_{A,J}$ are defined in Ref.~\cite{Helset:2018fgq}.
The vectors $t^{I}_{A}$ encode the isometries of the scalar field space manifold and are the Killing vectors of that manifold.
The SMEFT structure constants are
\begin{align}
	\f ABC = \tilde \epsilon^{A}_{\,\, BC}
\end{align}
defined in Ref.~\cite{Helset:2018fgq}.}

In the following, we assume that all particles are massless and that the gauge symmetry is linearly realized, so $t^{I}_{A}(v)=0$, where $v^{I}$ is the vacuum expectation value (VEV) of the scalar fields $\phi^I$. The results can be extended to not require these simplifying assumptions.
The kinetic term for the scalars depends on $h_{IJ}(\phi)$, which can be interpreted as a metric in field space for the scalars \cite{Meetz:1969as}. Similarly, the kinetic term for the gauge fields is multiplied by $g_{AB}(\phi)$ which depends on the scalars, is symmetric under $A \leftrightarrow B$, and transforms as an invariant tensor under action by the Killing vector $t^I_A$,
\begin{align}
\label{eq:LieMetricG}
g_{AB,I} \, t^I_C - \f DCA\ g_{DB} - \f DCB\ g_{AD} =0 \,.
\end{align}

We want to extend the notion of a field-space manifold to include the gauge fields, where $g_{AB}(\phi)$ will take center stage, and unify the scalar and gauge sectors, so Eqs.~\eqref{eq:LieMetricH} and~\eqref{eq:LieMetricG} will be components of a single equation.


We look at symmetries of the Lagrangian in Eq.~\eqref{eq:Lagr}, following closely the discussion in Ref.~\cite{Alonso:2016oah}.
The ungauged theory is invariant under an infinitesimal transformation of the scalar field,
\begin{align}
	\label{eq:scalarFieldTransform}
	\delta_{\theta} \phi^{I} = \theta^{A} t^{I}_{A} ,
\end{align}
where $\theta^{B}$ is an infinitesimal parameter and $t^{I}_{B}$ depends on the scalar field.
The partial derivative of the scalar field transforms as
\begin{align}
	\label{eq:scalarDerivativeTransform}
	\delta_{\theta} (\partial_{\mu} \phi^{I}) = \theta^{A} \, t^{I}_{A,J} (\partial_{\mu} \phi^{J}) .
\end{align}
We promote the global symmetry to a local symmetry by letting $\theta^{A}$ depend on space-time; $\theta^{A}(x)$.
We can define a covariant derivative which transforms as
\begin{align}
	\label{eq:covariantDerTransform}
	\delta_{\theta} (D_{\mu}\phi)^{I} = \theta^{A}(x) t^{I}_{A,J} (D_{\mu} \phi)^{J} \,,
\end{align}
by introducing a gauge field
\begin{align}
	\label{eq:covariantDerivative}
	(D_{\mu} \phi)^{I} = \partial_{\mu} \phi^{I} + A^{B}_{\mu} t^{I}_{B} \,,
\end{align}
as in Eq.~\eqref{eq:covD}.
The gauge field transforms as
\begin{align}
	\label{eq:gaugeTransform}
\delta_\theta A^B_\mu = & - \partial_\mu \theta^B - \f BCD \theta^C A^D_\mu\, ,
	\end{align}
and the field strength transforms as
\begin{align}
	\label{eq:fieldStrengthTransform}
	\delta_{\theta} F^{A}_{\mu\nu} =  - \f ABC \theta^{B} F^{C}_{\mu\nu} .
\end{align}

We observe the similarity between the transformations of the covariant derivative and the field strength in Eqs.~\eqref{eq:covariantDerTransform} and \eqref{eq:fieldStrengthTransform}. Also, in the Lagrangian, Eq.~\eqref{eq:Lagr}, the functions $h_{IJ}$ and $g_{AB}$ take on analogous roles. It is natural to ask whether there is a unified treatment of $h_{IJ}$ and $g_{AB}$, and a notion of Killing vectors that acts on both.

We combine the tensors into a unified metric
\begin{align}
\label{eq:metricVEV}
	\tilde g_{ij} = \begin{bmatrix}
		h_{IJ} & 0 \\
		0 &  -\eta_{\mu_A \mu_B} g_{AB} 
	\end{bmatrix} ,
\end{align}
where the index $i$ runs over both scalar indices $I$ and gauge indices $(A\mu_A)$.  $\tilde g$ can be viewed as the gauge-fixed metric in Feynman gauge. The covariant derivative $\tilde \nabla$ is defined using the metric $\tilde g$.
We define a combined Killing vector
\begin{align}
	\tilde t^{i}_{B} = \begin{bmatrix}
		t^{I}_{B} \\
		t^{(A\mu_A)}_{B} 
	\end{bmatrix} ,
\end{align}
that acts on scalars and gauge fields, where the gauge field piece is
\begin{align}
	t^{(A\mu_A)}_{B} &\equiv - \delta^{A}_{B} \partial^{\mu_A}-  \f ABC A^{C\mu_A} .
\end{align}
The Lie derivative of the combined metric becomes
\begin{align}
\label{eq:18}
	\tilde t^{k}_{C} \tilde g_{ij,k} + \tilde g_{ik} \tilde t^{k}_{C,j} + \tilde g_{kj} \tilde t^{k}_{C,i} = 0 .
\end{align}
This includes Eqs.~\eqref{eq:LieMetricH} and \eqref{eq:LieMetricG}. 
Furthermore, Eq.~\eqref{eq:gaugeTransform} takes the simple form
\begin{align}
	\delta_{\theta} A^{B\mu} = t^{(B\mu)}_{C} \theta^{C}
\end{align}
which is structurally similar to Eq.~\eqref{eq:scalarFieldTransform}.
We will see further justification for this grouping of metrics and Killing vectors shortly.


\section{Field-space curvature}

\subsection{Scalar field-space curvature}

We first look at the geometry of the submanifold of the scalar fields. The scalar fields take the role of coordinates in field space, with capital indices from the middle of the Latin alphabet. 
We identify the scalar field-space metric by looking at the symmetric matrix in front of $(\partial_{\mu} \phi)^{I} (\partial^{\mu} \phi)^{J}$ in the Lagrangian\footnote{The notion of a field-space metric can be extended to higher-derivative operators \cite{Cheung:2022vnd}.}
\begin{align}
	\mathcal{L} = \frac{1}{2} h_{IJ}(\phi) (\partial_{\mu} \phi)^{I} (\partial^{\mu}\phi)^{J} + \dots
\end{align}
The Christoffel symbol is defined with respect to the scalar field-space metric $h_{IJ}$,
\begin{align}
	\label{eq:scalarChristoffel}
	\Gamma^{I}_{JK} = \frac{1}{2} h^{IL} \left( h_{JL,K} + h_{LK,J} - h_{JK,L} \right) ,
\end{align}
where $h_{IJ,K} = \partial_{K} h_{IJ}$ and the Riemann curvature is
\begin{align}
\label{eq:scalarRiemann}
	&R_{IJKL} 
	\\ &= 
	h_{IM} \left( \partial_{K} \Gamma^{M}_{\,\,\, LJ} 
	- \partial_{L} \Gamma^{M}_{\,\,\, KJ} 
	+ \Gamma^{M}_{KN} \Gamma^{N}_{\,\,\, LJ} 
	- \Gamma^{M}_{LN} \Gamma^{N}_{\,\,\, KJ} \right) . \nonumber
\end{align}
The covariant derivative $\nabla_{I}$ is defined with the connection in Eq.~\eqref{eq:scalarChristoffel}. 
%
%

The field-space curvature is important; it shows up in scattering amplitudes, e.g., the scattering amplitude for four scalars depends on the Riemann curvature~\cite{Alonso:2015fsp,Alonso:2016oah},
\begin{align}
\mathcal{A}_{IJKL} = & R_{IJKL}\, s_{IK} + R_{IKJL}\, s_{IJ}\, , \nn
= &R_{IJKL}\, s_{13} + R_{IKJL}\, s_{12} \,,
\label{amp:R}
\end{align}
where $s_{ij}=(p_i+p_j)^2$ and $p_i$ is the incoming momentum of particle $i$, and numerical values for $i$ are the index position in the subscripts of $\mathcal{A}$.
The amplitude $\mathcal{A}_{IJKL}$ respects Bose symmetry because of the symmetry properties of $R_{IJKL}$, including the Bianchi identity $R_{IJKL}+R_{IKLJ}+R_{ILJK}=0$, and the kinematic identity $s_{IJ}+s_{IK}+s_{IL}=0$ which follows from momentum conservation and massless external particles.
The curvature also enters in the double-soft theorem for a general scalar effective field theory \cite{Cheung:2021yog}.
As is expected, the curvature transforms as a tensor under scalar field redefinitions. This is why scattering amplitudes depend on the Riemann curvature; it is field-basis invariant.\footnote{The Riemann curvature evaluated at the VEV is field-basis covariant rather than invariant. However, this subtle difference is irrelevant in physical observables like cross sections.}

\subsection{Curvature of the full manifold}

We now define the geometry including both scalars and gauge bosons. Recall that the natural indices for the coordinates are $i \in \{I, A\mu_{A} \}$ since the gauge bosons  have both a gauge group index and a space-time index. We also use the combined field-space metric Eq.~\eqref{eq:metricVEV}. 
All expressions are assumed to be evaluated at the vacuum unless indicated otherwise.
We use a bar or tilde notation to distinguish this metric and the definitions below from the corresponding objects in the scalar field space, such as Eqs.~\eqref{eq:scalarChristoffel} and \eqref{eq:scalarRiemann}.

We will also use the combined metric
\begin{align}
	\label{eq:fullMetric}
	\bar g_{ij} = \begin{bmatrix} 
		h_{IJ} & 0 \\
		0 &  g_{AB} \left( - \eta_{\mu_A \mu_B} + \frac{(p_{A})_{\mu_B} (p_{B})_{\mu_A}}{p_{A} \cdot p_{B}} \right)
	\end{bmatrix} .
\end{align}
The difference between Eqs.~\eqref{eq:metricVEV} and \eqref{eq:fullMetric} is whether or not a gauge-fixing term in Feynman gauge has been added. 
In the following calculations of geometric quantities we use Eq.~\eqref{eq:metricVEV} to define an inverse metric evaluated at the VEV and Eq.~\eqref{eq:fullMetric} for any derivatives of the metric, such as $
\bar g_{ij,k}$. The metric must be gauge-fixed to have a well-defined inverse metric, and Eq.~\eqref{eq:metricVEV} is the gauge-fixed metric in Feynman gauge. Derivatives of the inverse metric are evaluated using  $\partial_{k}(\bar g^{ij}) =  - \bar g^{ir} \bar g_{rs,k} \bar g^{sj} \rightarrow - \tilde g^{ir} \bar g_{rs,k} \tilde g^{sj}$.

Momentum-dependent metrics have been previously studied in Ref.~\cite{Cheung:2022vnd}. We take all momenta as incoming. All momentum-dependent objects have an overall momentum-conserving $\delta$-function, which we suppress.

The Christoffel symbols $\bar \Gamma^i_{jk}$ computed from $\bar g_{ij}$ are
\begin{align}
	\label{eq:scalarChristoffel2}
	\bar\Gamma^{i}_{jk} = \frac{1}{2} \tilde g^{il} \left( \bar g_{jl,k} + \bar g_{lk,j} - \bar g_{jk,l} \right) ,
\end{align}
where all index sums run over both the scalar indices $I$ and gauge indices $(A\mu_A)$. The components of $\bar \Gamma^i_{jk}$ are
\begin{align}
	\label{eq:ChristoffelIJK}
	\bar \Gamma^{I}_{JK} =& \Gamma^{I}_{JK} , \\
	\label{eq:ChristoffelAIJ}
	\bar \Gamma^{(A\mu_A)}_{IJ} =& \bar \Gamma^{I}_{(A\mu_A)J} =  \bar \Gamma^{(C\mu_C)}_{(A\mu_A)(B\mu_B)} = 0 , \\ \nn
	\label{eq:ChristoffelIAB}
	\bar \Gamma^{I}_{(A\mu_A)(B\mu_B)} =& \frac{1}{2} \nabla^{I} g_{AB} \left(\eta_{\mu_A\mu_B} - \frac{(p_A)_{\mu_B} (p_B)_{\mu_A}}{p_A \cdot p_B}  \right) \,, \\
	\bar \Gamma^{(A\mu_B)}_{(B\mu_B)I} =& g^{AC} \eta^{\mu_A \mu_C} h_{IJ} \bar \Gamma^{J}_{(B\mu_B)(C\mu_C)} \,.
\end{align}
The Riemann curvature, which is defined as 
\begin{align}
	\bar R_{ijkl} = 
	\bar g_{im} \left( \partial_{k} \bar \Gamma^{m}_{\,\,\, lj} 
	- \partial_{l} \bar \Gamma^{m}_{\,\,\, kj} 
	+ \bar \Gamma^{m}_{kn} \bar \Gamma^{n}_{\,\,\, lj} 
	- \bar \Gamma^{m}_{ln} \bar \Gamma^{n}_{\,\,\, kj} \right) ,
\end{align}
has components
\begin{widetext}
\begin{align}
	\label{eq:RiemannIJKL}
	\bar R_{IJKL} =& R_{IJKL} , \\
	\label{eq:RiemannAIJK}
	\bar R_{(A\mu_A)IJK} =& 0 ,
	\\
	\label{eq:RiemannAIBJ}
	\bar R_{(A\mu_A)IJ(B\mu_B)} =& 
	\frac{1}{2} \left( \nabla_{I} \nabla_{J} g_{AB} \right) \left( -\eta_{\mu_{A}\mu_{B}} + \frac{(p_A)_{\mu_B} (p_B)_{\mu_A}}{p_A \cdot p_B} \right)
	+ \frac{1}{4} \nabla_{J} g_{AC} g^{CD} \nabla_{I} g_{BD} 
	\biggl( \eta_{\mu_{A}\mu_{B}} 
	 \nonumber \\ & 
	- \frac{(p_{A})_{\mu_{B}} (p_{B} + p_{I})_{\mu_{A}}}{p_{A}\cdot (p_{B} + p_{I})} 
	- \frac{(p_{B})_{\mu_{A}} (p_{A} + p_{J})_{\mu_{B}}}{p_{B}\cdot (p_{A} + p_{J})}  + \frac{(p_B + p_I)_{\mu_{A}} (p_A + p_J)_{\mu_{B}} (p_A \cdot p_B)}{p_A\cdot(p_B + p_I )\ p_B\cdot (p_A + p_J)} \biggr) , \\
	\label{eq:RiemannABCI}
	\bar R_{(A\mu_A)(B\mu_B)(C\mu_C)I} =& 0 , \\
	\label{eq:RiemannABCD}
	\bar R_{(A\mu_A)(B\mu_B)(C\mu_C)(D\mu_D)} =& 
	- \frac{1}{4} \nabla_{I} g_{AC} \nabla^{I} g_{BD} \left( -\eta_{\mu_{A}\mu_{C}} + \frac{(p_A)_{\mu_C} (p_C)_{\mu_A}}{p_A \cdot p_C} \right)\left( -\eta_{\mu_{B}\mu_{D}} + \frac{(p_B)_{\mu_D} (p_D)_{\mu_B}}{p_B \cdot p_D} \right)
	\nonumber \\ &
	+ \frac{1}{4} \nabla_{I} g_{AD} \nabla^{I} g_{BC} \left( -\eta_{\mu_{A}\mu_{D}} + \frac{(p_A)_{\mu_D} (p_D)_{\mu_A}}{p_A \cdot p_D} \right)\left( -\eta_{\mu_{B}\mu_{C}} + \frac{(p_B)_{\mu_C} (p_C)_{\mu_B}}{p_B \cdot p_C} \right) .
\end{align}
\end{widetext}
The curvature for the extended scalar--gauge-field manifold obeys all the symmetry properties of the Riemann curvature, including the Bianchi identity
\begin{align}
	\bar R_{(A\mu_{A})(B\mu_{B}) IJ} + \bar R_{(A\mu_{A}) IJ(B\mu_{B})} + \bar R_{(A\mu_{A})J (B\mu_{B}) I}  = 0 .
	\label{Bianchi}
\end{align}
The virtue of these definitions becomes clear when we consider field redefinitions, gauge transformations, and scattering amplitudes.

\section{Field basis and gauge invariance}

Let's now consider properties of the new curvatures.
As one might expect from a field-space curvature, the extended curvatures are all invariant under scalar field redefinitions. Field redefinitions involving derivatives or gauge fields will in general mix operator structures with different number of derivatives, which complicates the picture of the invariances. Ref.~\cite{Cheung:2022vnd} showed that a generalized metric and descendant curvatures can still be constructed.

A second advantage is present when we have an external gauge index. The extended curvature can be contracted with the corresponding polarization vector for the gauge field,
\begin{align}\label{37}
	\bar R_{Ajkl} \equiv \epsilon^{\mu_{A}}_{A} \bar R_{(A\mu_A)jkl} .
\end{align}
Gauge invariance manifests itself for on-shell scattering amplitudes as Ward identities, with the replacement $\epsilon_{A} \rightarrow p_{A}$ giving a vanishing result.
Although the curvature is not a scattering amplitude, nor necessarily evaluated with on-shell kinematics, we can still ask about the replacement $\epsilon_{A}\rightarrow p_A$. Remarkably, we find that the curvatures in Eqs.~\eqref{eq:RiemannIJKL}--\eqref{eq:RiemannABCD} are gauge invariant:
\begin{align}
	p^{\mu_A}_A \bar R_{(A\mu_A) ijk} = 0 .
\end{align}
This holds even when the curvatures are not evaluated for on-shell kinematics. Also, descendant geometric objects such as $\bar \nabla_{i} \bar R_{jklm}$ will be gauge invariant in a similar fashion.



\section{Scattering Amplitudes}

Scattering amplitudes in scalar effective field theories depend on geometric invariants \cite{Volkov:1970aa}. This is natural since scattering amplitudes are invariant under field redefinitions. However, when we also include gauge fields in the theory, the original scalar curvatures are no longer sufficient to describe the scattering. In addition, the scattering amplitudes are gauge invariant. Therefore, we seek a generalization of the field-space geometry which includes both scalars and gauge fields and is gauge invariant. The candidates for such a geometry is already presented above. We will now see how it appears in scattering amplitudes.

The scattering amplitude for four scalars $\phi \phi \to \phi \phi$ (with a vanishing potential $V(\phi)$ for simplicity) is
\begin{widetext}
\begin{align}
	\label{eq:Amplitude4scalars}
	\cA_{IJKL} &= 
	R_{IJKL} s_{IK} + R_{IKJL} s_{IJ}  
	+ \frac{(t_{I;J} \cdot t_{K;L}) (s_{IL}-s_{IK})}{s_{IJ}} 
	+ \frac{(t_{I;K} \cdot t_{J;L}) (s_{IL}-s_{IJ})}{s_{IK}}
	+ \frac{(t_{I;L} \cdot t_{K;J}) (s_{IJ}-s_{IK})}{s_{IL}} , 
\end{align}
\end{widetext}
where $t_{IA;J} = \nabla_{J} t_{IA} = \nabla_J (h_{IK} t^K_A)= - t_{JA;I}$ and $t_{I;J} \cdot t_{K;L} = t_{IA;J} \, g^{AB} \, t_{KB;L}$. We see that the scattering amplitude depends on the field-space curvature in addition to the covariant derivative of the Killing vectors. The generalization of Eq.~\eqref{eq:Amplitude4scalars} including a potential $V(\phi)$ has additional terms involving covariant derivatives of $V(\phi)$.

In Eq.~\eqref{eq:Amplitude4scalars} we use the same indices for the scalars as in the Lagrangian in Eq.~\eqref{eq:Lagr}. This is a slight abuse of notation as the scalars appearing in the scattering amplitudes are the canonically-normalized mass eigenstates and not the generic flavor eigenstates in the Lagrangian. The translation between fields and states is performed by the tetrad which is derived from the metric $h_{IJ}$ \cite{Cheung:2021yog,Corbett:2019cwl}. Similarly, the gauge fields are canonically normalized and rotated to the mass-eigenstate basis using a tetrad coming from the metric $g_{AB}$.

It is possible to apply the geometry-kinematics map in Ref.~\cite{Cheung:2022vnd} to simplify the four-scalar amplitude Eq.~\eqref{eq:Amplitude4scalars} to
\begin{align}
	\label{eq:Amplitude4scalarsPrime}
	\cA_{IJKL} = & R^{\prime}_{IJKL}\, s_{13} + R^{\prime}_{IKJL}\, s_{12}   \,,
\end{align}
where the Riemann curvature $R^{\prime}$ depends on momentum and the Killing vectors,
\begin{align}\label{rprime}
R^{\prime}_{IJKL}  = & R_{IJKL} + \left( t_{I;L} \cdot t_{J;K} \frac{s_{IJ}-s_{IK}}{s_{IL}^2} - L \leftrightarrow K  \right) .
\end{align}
$R^{\prime}_{IJKL}$ has the same symmetry properties as $R_{IJKL}$. Eq.~\eqref{eq:Amplitude4scalarsPrime} has the same form as Eq.~\eqref{amp:R}.
In the following discussion, we will focus on the terms which do not depend on the Killing vectors $t^I_{A}$. The Killing vector terms require modifications to the formalism, such as $R_{IJKL} \to R^\prime_{IJKL}$ in Eq.~\eqref{rprime}, and a simple way to include such terms is still under investigation.

Another example where the geometry is present is the two scalar, two gauge boson scattering amplitude $\phi \phi \to W W$. Focusing on terms involving only the functions $g_{AB}$ and $h_{IJ}$ for simplicity, and ignoring other terms such as those involving the Killing vectors or the potential, the amplitude can be expressed in terms of two gauge-invariant structures \cite{Boels:2017gyc},
\begin{align}
	B_{1} =& (p_A \cdot p_B)(\epsilon_A \cdot \epsilon_B) - (\epsilon_A \cdot p_B)(\epsilon_B \cdot p_A)  , \\
	B_{2} =&   (\epsilon_A \cdot \epsilon_B)(p_A \cdot p_I) (p_B \cdot p_I) + (p_A \cdot p_B) (\epsilon_A \cdot p_I) (\epsilon_B \cdot p_I) \nn
	 -& (\epsilon_A \cdot p_B)(\epsilon_B \cdot p_I)(p_A \cdot p_I)
 -(\epsilon_A \cdot p_I)  (\epsilon_B \cdot p_A) (p_B \cdot p_I)  \,.
\end{align}
These terms of the scattering amplitude are
\begin{widetext}
\begin{align}
	\label{eq:Amplitude2scalar2gauge}
	\cA_{IJAB}=&
	\left(\nabla_{I} \nabla_{J} g_{AB} - \frac{1}{2} (\nabla_I g_{AC})g^{CD}(\nabla_{J} g_{BD}) - \frac{1}{2} (\nabla_J g_{AC})g^{CD}(\nabla_{I} g_{BD})  \right) B_{1} \nn &
	- \left(\frac{(\nabla_{I} g_{AC}) g^{CD} (\nabla_{J} g_{BD})}{ s_{IA} } + \frac{(\nabla_{J} g_{AC}) g^{CD} (\nabla_{I} g_{BD})}{ s_{JA} } \right) B_{2} + \dots
\end{align}
%
\end{widetext}
where the ellipsis denote terms which depend on other couplings than $g_{AB}$ and $h_{IJ}$. Again, the amplitude can be rewritten in terms of curvatures. The final expression is simply
\begin{align}
	\label{eq:amplitudeABIJ}
	\cA_{IJAB} = \bar R_{IJAB}\, s_{13} + \bar R_{IAJB}\, s_{12} ,
\end{align}
which is exactly the same form as the amplitude for four scalars in Eq.~\eqref{amp:R}, but using the curvature tensor $\bar R$ which is derived from the combined metric $\bar g$ in Eq.~\eqref{eq:fullMetric}. In fact, for all four-point amplitudes, the dependence on $g_{AB}$ and $h_{IJ}$ is simply 
\begin{align}
	\cA_{ijkl} = \bar R_{ijkl}\,  s_{13} + \bar R_{ikjl}\, s_{12} ,
\end{align}
for any combination of scalars and gauge fields.



\section{Geometry in the helicity basis}

Although all scattering amplitudes of scalars and gauge fields can be expressed in forms analogous to the non-linear sigma model \cite{Cheung:2022vnd}, the relevant curvature tensors will depend on kinematics and are in general  non-local. Any unphysical non-localities cancel when expanding the amplitude in terms of independent kinematic structures. We will see that the amplitudes can be unified using the combined kinematics-independent metric $\tilde g$ in Eq.~\eqref{eq:metricVEV}.

We focus on the couplings $h_{IJ}$ and $g_{AB}$ for the class of scattering amplitudes with two positive-helicity gauge bosons and additional scalars. 
The lowest-point scattering amplitude is
\begin{align}
	\cA^{++}_{IAB} =& \frac{1}{2} \nabla_{I} g_{AB}\, [AB]^2 .
\end{align}
where $[AB]$ is the spinor helicity product $[p_A p_B]$, etc.
%
This scattering amplitude describes the coupling of a Higgs boson to two photons or gluons.

Expanding the kinematic curvatures in Eq.~\eqref{eq:amplitudeABIJ} in terms of independent kinematic factors and fixing the helicities, we find that
\begin{widetext}
\begin{align}
	\label{eq:A4+-}
	\cA^{+-}_{IJAB}=&
	 \frac{1}{4} \left( (\nabla_{I} g_{AC}) g^{CD} (\nabla_{J} g_{BD}) \frac{ \langle IB \rangle^2 [IA]}{ \langle IA \rangle } + (\nabla_{J} g_{AC}) g^{CD} (\nabla_{I} g_{BD})  \frac{ \langle IB \rangle [IA]^2}{ [IB] }  \right) ,
	 \\
	\label{eq:A4++}
	\cA^{++}_{IJAB} =& \frac{1}{2} \tilde \nabla_{I} \nabla_{J} g_{AB} [AB]^2 ,
\end{align}
%
where 
\begin{align}
\tilde \nabla_{I} \nabla_{J} g_{AB}  = \nabla_{I} \nabla_{J} g_{AB} - \frac{1}{2} (\nabla_I g_{AC})g^{CD}(\nabla_{J} g_{BD}) - \frac{1}{2} (\nabla_J g_{AC})g^{CD}(\nabla_{I} g_{BD})   \,.
\end{align}
\end{widetext}
Note that the on-shell helicity amplitudes have only physical singularities.
The tilde covariant derivative $\tilde \nabla$ uses the connection defined from the metric in Eq.~\eqref{eq:metricVEV} with the $\eta_{\mu\nu}$ trivially stripped off. The tilde Christoffel symbols are the same as the bar Christoffel symbols in Eqs.~\eqref{eq:ChristoffelIJK}--\eqref{eq:ChristoffelIAB} after dropping the momentum-dependent terms. This notion of geometry is closely related to the construction in Ref.~\cite{Helset:2020yio}.

This structure continues for higher-point scattering amplitudes, e.g., for two positive-helicity gauge bosons and three scalars the amplitude is 
\begin{align}
	&\cA^{++}_{IJKAB} 
	\nonumber \\
	&= \frac{1}{2} \tilde \nabla_{(I} \tilde \nabla_{J} \nabla_{K)} g_{AB} [AB]^2 \nonumber
	\\ &
	+\left\{ \frac{(\nabla_{K} g_{AC})g^{CD}(\nabla_{J} g_{DE})g^{EF}(\nabla_{I}g_{FB})}{(p_A+p_K)^2 (p_B+p_I)^2}\left[\frac{1}{8} [A|p_K p_I|B]^2 \right.\right.
	\nonumber \\ &
	\left.\left.  \qquad-\frac{1}{6} (p_A\cdot p_K)(p_B\cdot p_I)[AB]^2 \right] +\textrm{perm}(I,J,K) \right\}
	\nonumber \\ &
	+ \frac{1}{6} \nabla^{L} g_{AB} [AB]^2 \frac{1}{s_{IJK}} \left[ s_{IJ}(R_{IJKL} - 2 R_{IKJL} )  
	\right. \nonumber \\ & \left.
	\qquad\qquad\qquad \qquad \qquad   + s_{IK}(R_{IKJL} - 2 R_{IJKL} ) 
	\right. \nonumber \\ & \left.
	\qquad\qquad\qquad \qquad \qquad   + s_{JK}(R_{IJKL} + R_{IKJL} ) 
	\right] .   
\end{align}
In general, the scattering amplitude for two positive-helicity gauge bosons and $n$ scalars takes the form
\begin{align}
	\cA^{++}_{I_{1}\cdots I_{n}AB} &= \frac{1}{2} \tilde \nabla_{(I_{1}} \cdots \tilde \nabla_{I_{n-1}} \nabla_{I_{n)}} g_{AB} [AB]^2 + \textrm{factorizable} 
\end{align}
where factorizable indicates terms which factorize to products of lower-point amplitudes for special kinematic configurations. 
The covariant derivative with the tilde connection, $\tilde \nabla$, plays a crucial role for the kinematic-independent geometry for scalars and gauge fields. 

This geometry in the helicity basis for scalars and gauge bosons is richer than the corresponding geometry for the scalars since it depends on several geometric quantities.
In the non-linear sigma model, the scattering amplitudes depend on $R_{IJKL}$, $\nabla_{I} R_{JKLM}$, etc., where the Riemann curvature $R_{IJKL}$ enters in the lowest-point non-vanishing amplitude. When including gauge bosons, new low-point amplitudes are relevant. The new structures in the scattering amplitudes are $\nabla_{I} g_{AB}$, $\tilde \nabla_{I} \nabla_{J} g_{AB}$, etc.



\section{Renormalization of invariants}

Since the new geometric structures are independent of the kinematics, they will also be present in higher-loop amplitudes and in the renormalization group equations.

An efficient method for extracting the anomalous dimension matrix for an effective field theory is to use scattering amplitudes and unitarity cuts \cite{Cheung:2015aba,Caron-Huot:2016cwu,Bern:2019wie,EliasMiro:2020tdv,Baratella:2020lzz,Jiang:2020mhe,Bern:2020ikv,AccettulliHuber:2021uoa}. At one loop, the relevant unitarity cuts are evaluated with two on-shell tree-level amplitudes, with two particles across the cut.
We can  set up renormalization group equations, not for the Wilson coefficients, but for the geometric structures. For example, 
\begin{widetext}
\begin{align}
	\frac{d R_{IKJL} }{d\log \mu} =& \frac{1}{16\pi^{2}} \frac{1}{2} \biggl\{ \frac{1}{4} \gamma_{\rm coll} R_{IKJL} 
		+  R_{I\;\;\;J\;}^{\;\;M\;\;N} \left[ V_{;(MNKL)}  + (t_{N;K} \cdot t_{M;L}) + (t_{N;L} \cdot t_{M;K}) \right]  \nn 
		& 
+ R_{IJ}^{\quad MN} \left[ - \frac{1}{6} (t_{N;M}\cdot t_{K;L}) + 4 (t_{N;L} \cdot t_{M;K}) \right]
+ R_{IK}^{\quad MN} \left[- \frac{1}{6} (t_{N;M}\cdot t_{J;L}) + 4 (t_{N;L} \cdot t_{M;J}) \right] + \dots 
			 \nn &  
		     + (I\leftrightarrow K, J\leftrightarrow L) 
		     - (J\leftrightarrow L) 
		     - (I\leftrightarrow K) 
	     \biggr\}
		     + \biggl\{ I\leftrightarrow J, K\leftrightarrow L \biggr\}
\end{align}
and 
\begin{align}
	\label{eq:RGEg}
	\frac{d (\tilde \nabla_{I} \nabla_{J} g_{AB}) }{d\log \mu} =& \frac{1}{16\pi^{2}} \biggl\{ \frac{1}{4} \gamma_{\rm coll} (\tilde \nabla_{I} \nabla_{J} g_{AB})
		+ \frac{1}{4} (\tilde \nabla^{K} \nabla^{L} g_{AB})\left[ V_{;(IJKL)} + 2 (t_{K;I} \cdot t_{L;J}) \right]
	\nn & \left.
	+(\tilde \nabla_{I} \nabla^{L} g_{AC}) g^{CD}  (4 t_{LB;K} \; t^{K}_{D;J} - 6t_{JB;K} \; t^{K}_{D;L}) + \dots
		\right. \nn
	& + (I\leftrightarrow J) + (A\leftrightarrow B)  + (I\leftrightarrow J, A\leftrightarrow B)  \biggr\}
\end{align}
\end{widetext}
where $\gamma_{\rm coll}$ is a collinear factor which only depends on the external particles, $\gamma_{\rm coll} = \gamma_I+\gamma_J+\gamma_K+\gamma_L $ and $\gamma_{\rm coll} = \gamma_I+\gamma_J+\gamma_A+\gamma_B $ in the two cases,
 and the ellipsis indicate terms which vanish for the Standard Model Effective Field Theory (SMEFT) in the unbroken phase.  For the SMEFT neglecting Yukawa couplings, $\gamma$ for the Higgs field and gauge bosons at one loop are\footnote{$\gamma_H$ in Feynman gauge is $\gamma_H=-(g_1^2+3g_2^2)/2$. The value in Eq.~\eqref{55} is with the gauge-fixing term used in Ref.~\cite{Helset:2022pde}.}
 \begin{align}
\gamma_H =&  - g_1^2 -3 g_2^2    \,, &
\gamma_A =& -  b_0 g^2\,,
\label{55}
\end{align}
where the one-loop $\beta$-function is $\beta(g) = - b_0 g^3/(16 \pi^2)$.
$\nabla_I g_{AB}$ vanishes for SMEFT in the unbroken phase, and so its renormalization group evolution is trivial.
Applying these equations to the SMEFT, we have verified that these expressions agree with the renormalization group equations for dimension-six operators in Refs.~\cite{Jenkins:2013zja,Jenkins:2013wua,Alonso:2013hga}. We have also used this geometric formalism to compute the renormalization group equations for dimension-eight bosonic operators in the SMEFT~\cite{Helset:2022pde}.



\section{Conclusions}

We have extended the notion of a field-space geometry to include both scalars and gauge bosons. This can be done in several ways---see Eqs.~\eqref{eq:metricVEV} and \eqref{eq:fullMetric}---and the resulting geometric invariants enter in the scattering amplitudes differently. Using Eq.~\eqref{eq:fullMetric}, the scattering amplitudes depend on a kinematic Riemann curvature, which is a special case of the geometry-kinematics duality of Ref.~\cite{Cheung:2022vnd}. What is remarkable about these kinematic curvatures is that they are invariant under both field redefinitions and gauge transformations. For helicity amplitudes, a closely related geometry given by Eq.~\eqref{eq:metricVEV}, which is independent of kinematics, plays an important role. This kinematic-independent geometry is is also present in the renormalization group equations; Eq.~\eqref{eq:RGEg}.

The methods discussed here have been used to compute the bosonic dimension-eight operator anomalous dimensions in SMEFT~\cite{Helset:2022pde}. We are investigating the extension of the method to  include fermions~\cite{Alonso:2017tdy,Finn:2020nvn,Assi:2023zid,Derda:2024jvo}, and also whether there is a simplified treatment of the Killing vector terms.


\section{Acknowledgments}

We thank Clifford Cheung, Julio Parra-Martinez, and Michael Trott for helpful discussions and comments on the draft. This work is supported in part by the U.S.\ Department of Energy (DOE) under award
number~DE-SC0009919.  A. H. is supported by the DOE under award number DE-SC0011632 and by the Walter Burke Institute for Theoretical Physics.

\bibliographystyle{utphys-modified}
\bibliography{bibliographyManifold.bib}

\end{document}